\documentclass[11pt,twoside]{article}

\usepackage{asp2014}

\aspSuppressVolSlug
\resetcounters

\begin{document}

\title{Dicover and access GAPS Time Series\\prototyping for interoperability}

\author{Marco Molinaro,$^1$ Serena Benatti,$^2$ Andrea Bignamini,$^1$ and Riccardo Claudi$^2$
\affil{$^1$INAF - Osservatorio Astronomico di Trieste, via G.B.~Tiepolo, 11 - 34143 Trieste, Italy; \email{marco.molinaro@inaf.it}}
\affil{$^2$INAF - Osservatorio Astronomico di Padova, vicolo Osservatorio, 5 - 35122 - Padova, Italy}}

\paperauthor{Marco Molinaro}{marco.molinaro@inaf.it}{0000-0001-5028-6041}{INAF}{OATs}{Trieste}{TS}{34143}{Italy}
\paperauthor{Serena Benatti}{serena.benatti@inaf.it}{0000-0002-4638-3495}{INAF}{OAPd}{Padova}{PD}{35122}{Italy}
\paperauthor{Andrea Bignamini}{andrea.bignamini@inaf.it}{ORCID_Or_Blank}{INAF}{OATs}{Trieste}{TS}{34143}{Italy}
\paperauthor{Riccardo Claudi}{riccardo.claudi@inaf.it}{ORCID_Or_Blank}{INAF}{OAPd}{Padova}{PD}{35122}{Italy}

\begin{abstract}
The GAPS (Global Architecture of Planetary Systems) project is a, mainly
Italian, effort for the comprehensive characterization of the
architectural properties of planetary systems as a function of the host
stars' characteristics by using radial velocities technique.
Since the beginning (2012) the project exploited the HARPS-N high
resolution optical spectrograph mounted at the 4-m class TNG
telescope in La Palma (Canary Islands). More recently, with the upgrade
of the TNG near-infrared spectrograph GIANO-B, obtained in the
framework of the GIARPS project, it has become possible to perform
simultaneous observations with these two instruments, providing
thus, at the same time, data both in the optical and in the
near-infrared range.
The large amount of data obtained in about 5 years of observations
provided various scientific outputs, and among them, time series of
radial velocity (RV) profiles of the investigated stellar systems.

This contribution shows the first steps undertaken to deploy the GAPS
Time Series as an interoperable resource within the VO framework
designed by the IVOA.
This effort has thus a double goal. On one side there's the aim at
making the time series data (from RV up to their originating spectra)
available to the general astrophysical community in an interoperable
way. On the other side, to provide use cases and a prototyping base
to the ongoing time domain priority effort at the IVOA level.
Time series dataset discovery, depicted through use cases and mapped
against the ObsCore model will be shown, highlighting
commonalities as well as missing metadata requirements.
Future development steps and criticalities, related also to the joint
discovery and access of datasets provided by both the spectrographs
operated side by side, will be summarized.
\end{abstract}

\section{Introduction}
The extrasolar planets search through the radial velocity (RV) method
takes advantage of
a large amount of spectroscopic datasets, coming from dedicated
observatories deployed in several countries.

A set of basic use cases has been developed to try to give first answers
to dataset discovery and access to time series for spectroscopic RV in
the exoplanets domain. This set of use cases (described in
Sec.~\ref{sec:ucs}), even if not exhaustive, nonetheless show that
probably a specific domain model is needed to be coupled with the
ObsCore table metadata \citep[IVOA ObsCore
Recommendation][]{std:OBSCORE11} and answer even some basic discovery
cases.

The use cases come from the experience in exoplanets discovery and
characterization gained by the GAPS project \citep{2013A&A...554A..28C}
and proceed from the results of the first years of the project
\citep[see, e.g.][]{2017arXiv170804166B}.

\section{Use Cases}
\label{sec:ucs}
Here we describe the use cases (UC) adopted for this contribution,
mapping case by case what fields of the ObsCore table are involved or
what extra tables and columns we used. Essentially we added, alongside
the mandatory and a few optional ObsCore fields, two dedicated tables:
\begin{description}
\item[exots] table to contain global metadata for the exoplanetary
system;
\item[exoplanets] to collect metadata for the exoplanets themselves.
\end{description}
This choice was made for two reasons: not to add new, domain specific,
fields to the general purpose ObsCore main table, and, for the
\textit{exoplanets} table, to allow the \mbox{1-to-N} relationship
between the host system and its exoplanets. The
\texttt{obs\_publisher\_did} ObsCore identifier field has been used to
glue the three tables together.

\begin{description}

\item[UC1] The first use case is to find all possible datasets that
contain radial velocity time series given, e.g., a region on the
celestial
sphere. This one is easily accomplished using the
\texttt{dataproduct\_type} field set to \textbf{timeseries} in
combination with the
\texttt{o\_ucd} (observable axis UCD) set to
\textbf{spect.dopplerVeloc.opt} (since we are currently dealing only
with optical data) and
relying on the positional search capabilities of the ObsCore model. For
future GIANO-B/GIARPS observations we'll need to add a
\textbf{.IR} leaf to the UCD branch or find an analogue solution to
support infrared observation derived products.

\item[UC2]ObsCore's use case 4.2 asks for constraints on the number of
points in the series, its time span and resolution. Our second
use case simply applies this to the exoplanets case. However, it seems
not clear what information the time axis parameters
(\texttt{t\_*} fields) should stand for, particularly the
\texttt{t\_resolution} field, given that the resolution of these time
series is
inherently uneven.

\item[UC3] The third use case deals with potentially detected
exoplanets. This is the first added metadatum, that we put in an
external
table (\textbf{exots}) linked to the ObsCore one. There are two fields
to deal with this: \texttt{candidates} and \texttt{confirmed} (see Table
\ref{tab:tstable}) to filter for candidate or confirmed planets in an host
system.

\item[UC4] This use case is meant to identify what discovery method is
used for the exoplanets whose time series are searched for. We
set also this parameter \texttt{method} in the \textbf{exots} new table,
setting it to textbf{RVspectroscopy}, but considering that this field
should have values coming from a controlled vocabulary of discovery
methods (transit, direct-imaging, astrometry, \ldots). The use case
may also be solved adapting, again, the \texttt{o\_ucd} to express the
discovery method, but it would be confusing in the usage, because
it would mean mapping two distinct concepts in the same field.

\begin{table}[!ht]
\caption{\textbf{exots}: additional table to deal with exoplanets global
parameters as explained in the presented use cases (see listing in
Sec.~\ref{sec:ucs}).}
\label{tab:tstable}
\begin{center}
{\small
\begin{tabular}{l|l}  
\tableline
\noalign{\smallskip}
column name & description \\
\noalign{\smallskip}
\tableline
\noalign{\smallskip}
\texttt{obs\_publisher\_did} & dataset identifier reference ley to the
obscore table \\
\texttt{candidates} & number of candidate and confirmed planets in the
stellar system \\
\texttt{confirmed} & number of confirmed planets in the
stellar system \\
\texttt{method} & discovery method used to detect the exoplanets
(vocabulary based) \\
\texttt{host\_activity} & host star activity index \\
\texttt{host\_mass} & host star mass \\
\texttt{host\_type} & host star spectral type \\
\texttt{host\_metallicity} & host star metallicity \\
\texttt{host\_age} & host star age \\
\texttt{systemic\_RV} & systemic radial velocity calculated for the
system \\
\texttt{updated} & last time the time series has been workd on\\
\noalign{\smallskip}
\tableline
\end{tabular}
}
\end{center}
\end{table}

\begin{table}[!ht]
\caption{\textbf{exoplanets}: additional table to deal with exoplanets
parameter values (see UC6 listing in Sec.~\ref{sec:ucs}). Differently to
the case of the \textbf{exots} table, here the
\texttt{obs\_publisher\_did} defines a \mbox{1-to-N} relation to the
\textbf{exots} tables.}
\label{tab:exoplanets}
\begin{center}
{\small
\begin{tabular}{l|l}  
\tableline
\noalign{\smallskip}
column name & description \\
\noalign{\smallskip}
\tableline
\noalign{\smallskip}
\texttt{obs\_publisher\_did} & dataset identifier reference ley to the
exots table \\
\texttt{planet\_id} & exoplanets identifier in the system (b, c, d,
\ldots) \\
\texttt{msini} & exoplanet minimum mass parameter \\
\texttt{period} & period \\
\texttt{eccentricity} & orbital eccentricity \\
\texttt{RVsemiamplitude} & radial velocity semi-amplitude \\
\texttt{t0} & time at periastron (RV) or central transit time (transit)
\\
\texttt{omega} & periastron longitude \\
\noalign{\smallskip}
\tableline
\end{tabular}
}
\end{center}
\end{table}

\item[UC5] Also host star characterization is important, so we have a
use case that tries to solve discovery for stars having, e.g., low
stellar activity (currently GAPS hosts are all low activity stars, but
this may change in the future with adoption of IR spectroscopy) or a
specific spectral type or similar. See the \texttt{host\_*} fields in
Table~\ref{tab:tstable} for an example set of star characteristics that are
used.

\item[UC6] Planets parameter values and their ranges define another use
case for filtering time series of exoplanets. Since this requires
multiple values for each system, another custom table
(\textbf{exoplanets}, Table~\ref{tab:exoplanets}) has been added to
include planets'
data like mass and orbital coefficients.

\item[UC7] This use case is not discussed here, but it mainly focuses on
linking the time series points to their originating spectra, thus
providing provenance access the datasets used to build the time series.
The idea is to use the Datalink (IVOA Recommendation, see
\citet{std:DataLink}) for the \texttt{access\_url} and
\texttt{access\_format} ObsCore fields.

\item[UC8] Also the last use case is not discussed here. It asks to find
time series of photometry of the host system, to use in combination with
spectral data in the exoplanets identification. Standard ObsCore should
be able to solve it.
\end{description}

\section{Conclusions}
Simple discovery and access use cases for exoplanets time series do not
look a difficult task adopting an IVOA ObsCore solution (see use
cases 1 and 2 in Sec.~\ref{sec:ucs}). However some changes may be needed
if we want specific discovery scenarios to work, like use
cases from 3 to 6. Moreover, some information, useful when dealing with
spectroscopic RV time series, can be set using ObsCore fields
(\texttt{s\_fov} and the \texttt{em\_*} fields) but may be misleading
since they refer to the spectra from which the time series originate,
rather than being a direct description of the series points. 

Also, \texttt{t\_*} fields (the time axis characterisation ones), that play a specific
role in the scenario, are quite confusing or misleading for time series,
where the concepts of resolution, exposure time and start/stop of
an observation are quite different from those of a single observation.

The solution here presented, using a couple more tables referencing
 the ObsCore through the \texttt{obs\_publisher\_did} can be a solution
(given we use a helper solution setting the
\texttt{dataproduct\_subtype} to \textbf{RV:optical} to allow for mixed
ObsCore content).
It would probably be better if a simple model for time series is
developed, to which the depicted tables (modified, updated, generalized)
follow as flat views to be connected to the ObsCore main table. 

The two final uses cases, not solved within this contribution, should
not present criticalities, because use case 7 fits the IVOA
Datalink goals, while use case 8 should be already solved in ObsCore.

%

\acknowledgements This contribution benefits funding from the ASTERICS
pro-ject, supported by the European Commission Framework Programme
Horizon 2020 Research and Innovation action under grant agreement n.
653477.

\bibliography{P11-55}  

\end{document}